\documentclass[sigconf]{acmart}
\usepackage{algorithmicx}
\usepackage{url}
\usepackage{bbm}
\usepackage{bm}
\usepackage{subcaption}
\usepackage[export]{adjustbox}
\usepackage{amsmath}
\usepackage{mathtools}
\usepackage{multirow}
\usepackage{makecell}
\usepackage{amsthm}
\usepackage{algpseudocode}
\usepackage{balance}
\usepackage{xcolor}
\usepackage[linesnumbered,ruled,vlined]{algorithm2e}

\SetCommentSty{mycommfont}

\SetKwInput{KwInput}{Input}                
\SetKwInput{KwOutput}{Output}           
\SetKwInput{KwInit}{Initialization}
\newtheorem{definition}{Definition}
\DeclareMathOperator*{\argmax}{argmax}

\AtBeginDocument{%
  \providecommand\BibTeX{{%
    \normalfont B\kern-0.5em{\scshape i\kern-0.25em b}\kern-0.8em\TeX}}}

\copyrightyear{2024}
\acmYear{2024}
\setcopyright{acmlicensed}
\acmConference[SIGIR '24] {Proceedings of the 47th International ACM SIGIR Conference on Research and Development in Information Retrieval}{July 14--18, 2024}{Washington, DC, USA.}
\acmBooktitle{Proceedings of the 47th International ACM SIGIR Conference on Research and Development in Information Retrieval (SIGIR '24), July 14--18, 2024, Washington, DC, USA}
\acmISBN{979-8-4007-0431-4/24/07}
\acmDOI{10.1145/3626772.3661357}

\begin{document}
\title{LLM-Ensemble: Optimal Large Language Model Ensemble Method for E-commerce Product Attribute Value Extraction}

\author{Chenhao Fang}
\authornote{Both authors contributed equally to this research.}
\authornote{Work done while at Walmart.}
\affiliation{%
  \institution{University of Wisconsin-Madison}
  \state{Wisconsin}
  \country{USA}
}
\email{chenhao.fang@outlook.com}

\author{Xiaohan Li}
\authornotemark[1]
\affiliation{%
  \institution{Walmart Global Tech}
  \city{Sunnyvale}
  \state{California}
  \country{USA}}
\email{xiaohan.li@walmart.com}

\author{Zezhong Fan}
\affiliation{%
  \institution{Walmart Global Tech}
  \city{Sunnyvale}
  \state{California}
  \country{USA}}
\email{zezhong.fan@walmart.com}

\author{Jianpeng Xu}
\affiliation{%
  \institution{Walmart Global Tech}
  \city{Sunnyvale}
  \state{California}
  \country{USA}}
\email{jianpeng.xu@walmart.com}

\author{Kaushiki Nag}
\affiliation{%
  \institution{Walmart Global Tech}
  \city{Sunnyvale}
  \state{California}
  \country{USA}}
\email{kaushiki.nag@walmart.com}

\author{Evren Korpeoglu}
\affiliation{%
  \institution{Walmart Global Tech}
  \city{Sunnyvale}
  \state{California}
  \country{USA}}
\email{ekorpeoglu@walmart.com}

\author{Sushant Kumar}
\affiliation{%
  \institution{Walmart Global Tech}
  \city{Sunnyvale}
  \state{California}
  \country{USA}}
\email{sushant.kumar@walmart.com}

\author{Kannan Achan}
\affiliation{%
  \institution{Walmart Global Tech}
  \city{Sunnyvale}
  \state{California}
  \country{USA}}
\email{kannan.achan@walmart.com}
\renewcommand{\shortauthors}{Chenhao Fang et al.}

\begin{abstract}
    Product attribute value extraction is a pivotal component in Natural Language Processing (NLP) and the contemporary e-commerce industry. The provision of precise product attribute values is fundamental in ensuring high-quality recommendations and enhancing customer satisfaction. The recently emerging Large Language Models (LLMs) have demonstrated state-of-the-art performance in numerous attribute extraction tasks, without the need for domain-specific training data. Nevertheless, varying strengths and weaknesses are exhibited by different LLMs due to the diversity in data, architectures, and hyperparameters. This variation makes them complementary to each other, with no single LLM dominating all others. Considering the diverse strengths and weaknesses of LLMs, it becomes necessary to develop an ensemble method that leverages their complementary potentials.  

In this paper, we propose a novel algorithm called LLM-ensemble to ensemble different LLMs' outputs for attribute value extraction. We iteratively learn the weights for different LLMs to aggregate the labels with weights to predict the final attribute value. Not only can our proposed method be proven theoretically optimal, but it also ensures efficient computation, fast convergence, and safe deployment. We have also conducted extensive experiments with various state-of-the-art LLMs on Walmart's internal data. Our offline metrics demonstrate that the LLM-ensemble method outperforms all the state-of-the-art single LLMs on Walmart's internal dataset. This method has been launched in several production models, leading to improved Gross Merchandise Volume (GMV), Click-Through Rate (CTR), Conversion Rate (CVR), and Add-to-Cart Rate (ATC). 

\end{abstract}



\maketitle

\section{Introduction}
With the development of Natural Language Processing (NLP) techniques and their applications in the e-commerce industry, the extraction of accurate product attribute values with NLP plays a critical role~\cite{ghani2006text, more2016attribute, rezk2019accurate}. The quality and relevance of product recommendations, crucial to enhancing customer satisfaction, are heavily reliant on the precision of these attributes. However, a significant challenge faced by e-commerce platforms is the lack of access to precise attribute data, leading to less accurate recommendations. Existing methods for attribute extraction are evaluated on the high-quality datasets from other platforms~\cite{zheng2018opentag, xu2019scaling, yan2021adatag}; however, these methods often falter when applied to Walmart's internal datasets, resulting in less accurate extractions.

Recent advancements in the field of NLP have seen the emergence of Large Language Models (LLMs), which have shown exceptional performance in a variety of NLP tasks, including attribute value exaction, notably without the necessity for domain-specific training data. These models, including Llama~\cite{touvron2023llama}, GPT~\cite{brown2020language}, and PaLM~\cite{chowdhery2023palm}, have revolutionized the way attribute value extraction is approached, offering a new level of efficiency and accuracy. Brinkmann et al.~\cite{brinkmann2023product} have also demonstrated using LLMs can also significantly improve the accuracy of the product attribute value extraction and achieve state-of-the-art performance. Despite their effectiveness, these LLMs exhibit distinct strengths and weaknesses due to differences in their underlying data sources, architectural designs, and hyperparameters. This diversity results in a scenario where no single LLM is universally superior across all tasks.

In light of these variations, it becomes essential to explore ensemble methods that leverage the complementary strengths of different LLMs. Ensemble methods in machine learning~\cite{dietterich2000ensemble} can enhance the robustness, generalization, and accuracy of attribute extraction by aggregating the unique contributions of each model. They are particularly effective in mitigating biases, errors, and uncertainties inherent in individual models, thereby aligning the results more closely with human judgment and preferences. At Walmart, we offer hundreds of millions of items spanning numerous categories. Due to variations in the training datasets, different LLMs demonstrate varying levels of performance across these item categories. Therefore, experimenting with different LLMs for each category is not only costly but also time-intensive. This necessitates the development of an efficient strategy to seamlessly integrate multiple LLMs, enhancing overall effectiveness and efficiency.

Recently, there have been some papers on LLM fusion, as highlighted by works in~\cite{jiang2023llm, wan2024knowledge}. These studies primarily aim to enhance text generation capabilities in areas such as code generation and reasoning. However, when it comes to extracting product attribute values, the requirements are notably different. In this context, LLMs are tasked with generating concise outputs, such as a single word or a short phrase, representing the attribute values of products. To address this unique challenge, we draw inspiration from crowdsourcing techniques, treating each LLM as an individual worker. Through a process of voting and iterative refinement of predictions, our algorithm assigns weights to each LLM, calibrating their influence based on their demonstrated accuracy in specific tasks.

This paper introduces a novel algorithm called LLM-ensemble designed to ensemble the outputs of various LLMs for the purpose of attribute extraction. At its core, our approach is based on the Dawid-Skene Model~\cite{dawid1979maximum}, a structured latent variable model, to iteratively learn and assign weights to different LLM outputs. Our method is not only theoretically optimal but also boasts efficient computation, and rapid convergence, and ensures safe deployment. We validate our approach through extensive experimentation with leading LLMs on Walmart's internally labeled data. The results from these experiments clearly demonstrate that our LLM-ensemble method surpasses the performance of any single state-of-the-art LLM on Walmart's dataset.

Furthermore, we have successfully deployed this method, generating millions of highly accurate "age" and "gender" attribute labels for items in Walmart. The deployment of this method in various production models has yielded tangible benefits, including significant improvements in Gross Merchandise Volume (GMV), Click-Through Rate (CTR), Conversion Rate (CVR), and Add-to-Cart rate (ATC). The integration of this algorithm into Walmart's e-commerce platform marks a significant advancement in the field of NLP and e-commerce, setting a new standard for attribute extraction and recommendation quality.
\section{Related Works}
\subsection{Product Attribute Value Extraction}
Early research on attribute value extraction relied on domain-specific rules to identify attribute-value pairs from product descriptions, as indicated in~\cite{zhang2009framework, vandic2012faceted}. However, the initial learning-based approaches required substantial feature engineering and struggled to adapt to previously unseen attributes and values~\cite{ghani2006text, putthividhya2011bootstrapped, wong2009scalable}. Recently, some studies~\cite{kozareva2016recognizing, zheng2018opentag} have shifted towards employing BiLSTM-CRF architectures for tagging attribute values in product titles. OpenTag~\cite{zheng2018opentag} leverages a BiLSTM-CRF model enhanced by active learning. Further innovations include SU-OpenTag~\cite{xu2019scaling}, which extends OpenTag by incorporating both a target attribute and the product title into a pre-trained language model. AdaTag~\cite{yan2021adatag} introduces a combination of a language model and a mixture-of-experts module for attribute value extraction, while TXtract~\cite{karamanolakis2020txtract} integrates a product taxonomy into its model. Approaches like AVEQA~\cite{wang2020learning} and MAVEQA~\cite{yang2022mave} frame attribute value extraction as a question-answering task, utilizing diverse pre-trained language models to process the target attribute, product category, and title. OA-Mine~\cite{zhang2022oa} explores the mining of unknown attribute values and attributes using a language model. More recent research has explored soft prompt tuning to fine-tune a minimal number of trainable parameters within a language model~\cite{blume2023generative, yang2023mixpave}. Additionally, Brinkmann et al.~\cite{brinkmann2023product} have demonstrated the extraction of product attribute values using Large Language Models (LLMs) like GPT-3.5 and GPT-4, highlighting the ongoing evolution in this field. Zou et al.~\cite{zou2024implicitave, zou2024eiven} propose ImplicitAVE and EIVEN to extract implicit attribute values with multimodal large language models.

\subsection{Feature Extraction with LLMs}
LLMs often outperform other models in zero-shot learning tasks and exhibit greater robustness when faced with unseen examples~\cite{brown2020language}. This superior performance is attributed to their extensive pre-training on vast text corpora and the emergent abilities~\cite{wei2022emergent} arising from their substantial model sizes. LLMs have demonstrated effectiveness across various application domains, particularly in information extraction tasks. For instance, Wang et al.~\cite{wang2023code4struct} and Parekh et al.~\cite{parekh2023geneva} have utilized OpenAI's LLMs for extracting structured event data from unstructured text sources. Agrawal et al.~\cite{agrawal2022large} have applied InstructGPT, leveraging zero-shot and few-shot learning prompts, to extract information from clinical notes efficiently. Similarly, Chen et al.~\cite{chen2023knowledge} have used LLMs to identify relationships between products in the e-commerce sector. Maragheh et al.~\cite{maragheh2023llm} extract keywords for products that are inferred from their textual data. Furthermore, LLMs have been employed to rank items by extracting features directly from prompts~\cite{hou2023large}. Despite these advances, current methodologies primarily focus on employing a single LLM, overlooking the potential benefits of ensemble approaches that could combine multiple LLMs to achieve enhanced results.
\begin{figure*}
    \centering
    \includegraphics[width=0.8\textwidth]{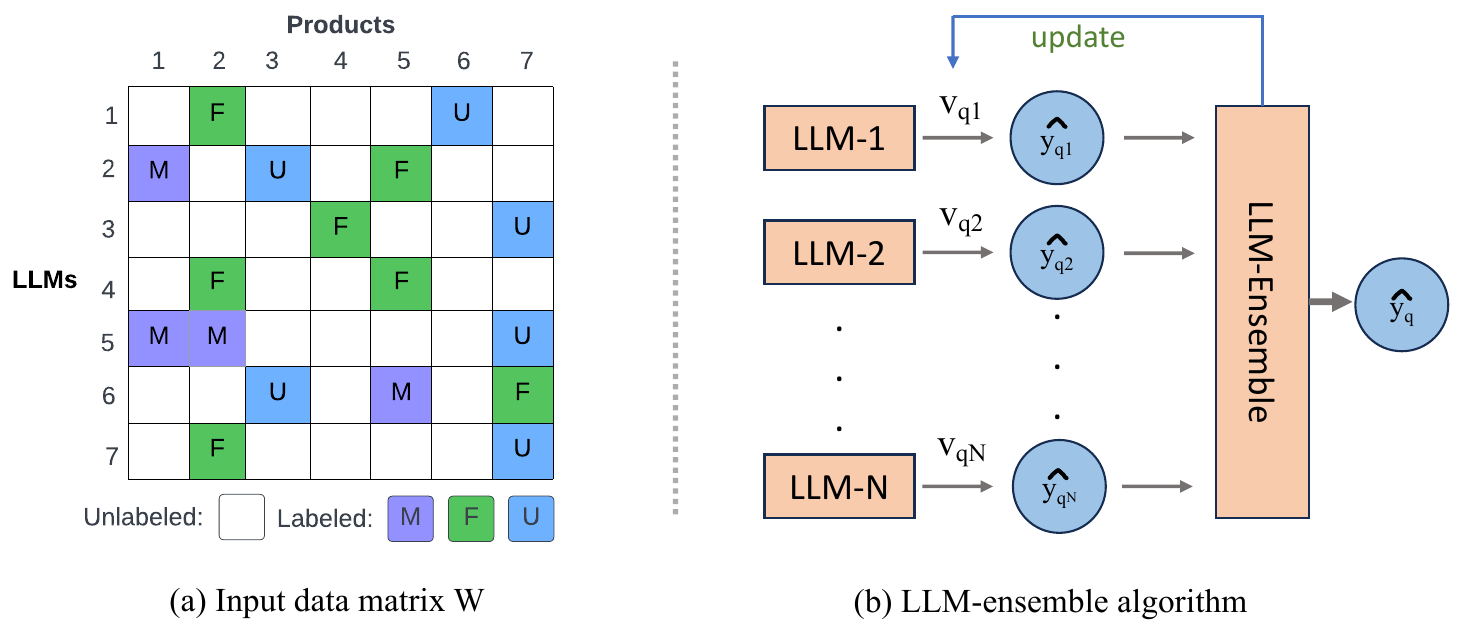}
    \caption{(a) The input data matrix $W$. We take the attribute "gender" as an example, and its labels are "Male" (M), "Female" (F), and "Unisex" (U). (b) The illustration of LLM-ensemble procedures. To learn the label of a product for attribute $q$, we have $N$ LLMs as inputs to the LLM-Ensemble algorithm. After several rounds of iteration, the algorithm generates the weights for each LLM and aggregates the labels with weights to predict the final label $\hat{y}_q$.}
    \label{fig:model}
\end{figure*}

\section{Methodology}
In this section, we introduce how the proposed LLM-ensemble algorithm utilizes multiple LLMs to obtain better predictive performance on the attributes in e-commerce. Motivated from ~\cite{li2014error, tian2015max, chen2022label}, we leverage the crowdsourcing techniques to ensemble multiple LLMs to achieve enhanced results.

\subsection{Problem Definition}
Our goal is to extract specific attribute values from unstructured text data, such as product profiles that include titles and descriptions, based on a set of pre-defined target attributes (e.g., gender, age, style). We aim to identify attribute values that have not been previously labeled. To achieve this, we have predetermined a set of applicable attribute values for products within the domain. For example, given the inputs,
\begin{itemize}
    \item target attributes: gender, age, and size
    \item product title: "Garanimals Toddler Girl Short Sleeve Graphic T-Shirt, Sizes 18M-5T"
    \item product description: "Bring an instant smile to her face with this colorful Graphic T-shirt from Garanimals. Cute and comfortable in a soft knit fabric $\cdots$"
\end{itemize}
Based on this unstructured text data, we want to extract 'female' (gender), 'child' (age), and 't-shirt' (style) as the corresponding values as
output from LLMs. Formally, our problem is defined as

\begin{definition}[Attribute Value Extraction with LLM ensemble.]
\label{definition}
Given a set of products $\mathcal{P}$, we utilize their unstructured text data $\mathcal{T} = \{t_1, \cdots, t_p: p \in \mathcal{P}\}$ and a set of attributes $\mathcal{Q} = \{q_1, \cdots, q_m \}$ to extract the corresponding attribute-values $\mathcal{V}_p = \{v_{p,1}, \cdots, v_{p,q}\}$ for $p \in \mathcal{P}$ and $q \in \mathcal{Q}$. $m$ is the number of the pre-defined attributes. The attribute values in $\mathcal{V}_p$ are selected from $\mathcal{L}_q$, which is the set of pre-defined labels.
\end{definition}


\subsection{LLM Ensemble}
As an example of ensemble learning, we assume that a set of LLMs is assigned to perform a labeling task extracting the specific values from the product attributes. In the following parts, we use "label" to represent the attribute values as they are selected from a limit set. Based on the Dawid-Skene Model~\cite{dawid1979maximum}, a structured latent variable model, our algorithm iteratively learns and assigns weights to different LLM outputs. We assume that there is a set of LLMs $\mathcal{N}$ and $|\mathcal{P}|$ products for the labeling task of attribute $q$ with $|\mathcal{L}_q|$ label classes.  The extended label set includes missing values represented by 0, which is defined as $\bar{\mathcal{L}}_q = \mathcal{L}_q \cup \{0\}$.

Subsequently, we take $y_{qp}$ as the ground-truth label of the $p$-th product for attribute $q$, and $\hat{y}_{qp}$ as the predicted label for the $p$-th product and attribute $q$ by an LLM. The input data matrix is denoted by $W \in \bar{\mathcal{L}}_q^{N \times P}$ where $N = |\mathcal{N}|$ and $P = |\mathcal{P}|$. $W_{qij}$ is the label provided by $i$-th LLM to the $j$-th product for attribute $q$. The missing corresponding label is represented by 0, which means the
$i$-th LLM hasn't labeled the $j$-th product yet. We introduce the indicator matrix $T = (T_{qij})_{N \times P}$ for attribute $q$, where $T_{qij} = 1$ indicates that entry $(i, j)$ is observed, and $T_{qij} = 0$ indicates entry $(i, j)$ is unobserved. Please note that $W$ and $T$ are observed together. To learn the weight $v_i$ for the $i$-th LLM, we illustrate our LLM-ensemble algorithm in Algorithm~\ref{alg:ensemble}.

Based on the findings from ~\cite{li2014error}, we prove that our algorithm is theoretically optimal to ensemble multiple LLMs. Under our problem settings, the oracle Maximum A Posteriori (MAP) rule~\cite{li2014error} approximately optimizes the upper bound on the mean error rate of weighted majority voting of LLMs. Our iterative weighted LLM-ensemble algorithms further optimize the error rate bound and approximate the oracle MAP rule. Therefore, with our LLM-ensemble method, we can assign higher weights to the "superior" LLMs, while mitigating the influence of the "spammers" (referring to LLMs whose accuracy is comparable to random guessing).

\begin{algorithm}[!ht]
\DontPrintSemicolon
  
  \KwInput{Number of LLMs = $N$; Number of products= $P$; Attribute $p$; input data matrix: $W \in \bar{\mathcal{L}}_q^{N \times P}$}
  \KwOutput{The predicted attribute values for attribute $q$: $\{\hat{y}_{q1}, \cdots, \hat{y}_{qP}\} $}
  \KwInit{$v_{qi} = 1, \forall i \in \mathcal{N}; T_{qij} = I(W_{qij} \ne 0), \forall i \in \mathcal{N}, \forall j \in \mathcal{P}$\tcp*{I is the indicator matrix}} 
  \tcc{loop the following steps to learn weights for LLMs}
   \While{not converges \textbf{or} reaches maximum iterations}
   {
   	$\hat{y}_{qj}\leftarrow \argmax \sum_{i=1}^N v_{qi}I(W_{qij}=k), \forall j \in \mathcal{P}$  \;
        $\hat{\alpha}_{qi} \leftarrow \frac{\sum_{j=1}^PI(W_{qij} = \hat{y}_{qj})}{\sum_{j=1}^PT_{qij}}, \forall i \in \mathcal{N}$ \;
        $v_{qi} \leftarrow L \hat{\alpha}_{qi} -1, \forall i \in \mathcal{N}$
        
   }
\textbf{return} the predictions $\{\hat{y}_{qj}\}_{j\in \mathcal{P}}$ by $\{\hat{y}_{qj}\} = \argmax_{k\in \mathcal{L}_q}\sum_{i=1}^N v_{qi}I(W_{ij}=k)$
\caption{LLM-ensemble algorithm for product attribute-value extraction of attribute $q$.}
\label{alg:ensemble}
\end{algorithm}

\section{Experiments}
We conduct two experiments to verify the effectiveness of our method. The first is comparison experiments, in which we compare the LLM-ensemble with all single LLMs as well as traditional methods to extract attribute values. In the second experiment, we show the results of the A/B test on the similar item recommendation model in Walmart.

\subsection{Comparison Experiments}
Our proposed LLM-ensemble is compared with its base LLMs: Llama2-13B, Llama2-70B~\cite{touvron2023llama}, PaLM-2~\cite{chowdhery2023palm}, GPT-3.5, GPT-4~\cite{brown2020language}. We also include logistic regression~\cite{hosmer2013applied} and our internal rule-based method in the baseline models. The datasets we employ are Walmart-Age and Walmart-Gender, which contain products sensitive to the ages or genders of customers\footnote{Due to the privacy policy in Walmart, we only disclose these two attributes. The actual product in practice has more attributes involved.}. Each dataset has 20K items for offline evaluation. The ground-truth label is created by the crowdsourcing results.

The experiment results are shown in Table~\ref{tab:score}. From this table, we can find that our proposed LLM-ensemble achieves the best performance compared to all other baseline models, which demonstrates the effectiveness of the LLM ensemble algorithm. Moreover, the logistic regression and rule-based method are much worse than LLM-based model, which means LLMs' emergent abilities can dramatically improve the accuracy of the product attribute value extraction.

\begin{table}[]
\centering
\begin{tabular}{l|l|l}
\toprule
Models                  & \multicolumn{1}{c|}{Walmart-Age} & {Walmart-Gender} \\ \midrule
Logistic regression           & 0.653                           & 0.681      \\ \hline
Rule-based method         & 0.710                           & 0.759      \\ \hline
Llama2-13B   & 0.753                           & 0.798      \\ \hline
Llama2-70B  & 0.887                           & 0.910      \\ \hline
PaLM-2  & 0.875                           & 0.894      \\ \hline
GPT-3.5  & 0.911                           & 0.933      \\ \hline
\underline{GPT-4}  & \underline{0.934}   & \underline{0.952}      \\ \hline
\textbf{LLM-ensemble}  & \textbf{0.956}    & \textbf{0.979}      \\ \hline
Improvement  & 2.36\%  & 2.76\%  \\ \bottomrule
\end{tabular}

\caption{Comparison experiments of different models on the prediction accuracy. The underlined model is the second-best one and the \textbf{bold} model is the best of all models.}
\label{tab:score}
\end{table}

\subsection{A/B test on Walmart Recommendation Model}
Having demonstrated state-of-the-art performance in our offline experiments, we've chosen to advance to an online A/B test. Within these trials, we apply the age and gender labels as top-layer filtration on the similar item recommendation model, which is one of the recommendation models in Walmart, to filter out misclassified products. 
By employing our method,  the recommendation model can enhance user engagement and increase conversion rates by ensuring the relevance of recommendation results. We implement this top-layer filtration across over 200 product categories on Walmart's e-commerce platform.

Table ~\ref{online-experiments} shows the outcomes of the online experiment, demonstrating a statistically significant enhancement across various key metrics in e-commerce, including Gross merchandise volume (GMV), Click-Through Rate (CTR), Conversion Rate (CVR) and Add-to-Cart Rate (ATC). 
These findings strongly support the effectiveness of the LLM-ensemble methodology, emphasizing its importance and potential impact on e-commerce applications. As a result of the success of the A/B test, we have now launched this feature on Walmart's online platform.

\begin{table}[h]
  \caption{Online experiment results}
  \label{online-experiments}
  \centering
  \begin{tabular}{l|l|l}
    \toprule            
    Metrics  & \multicolumn{1}{c|}{Percentage Lift} & {P-Value}  \\
    \midrule
    GMV & 0.38\%  & 0.039  \\ \hline
    CTR & 2.16\%  & 0.028   \\ \hline
    CVR & 0.26\%  & 0.043\\ \hline
    ATC  & 1.42\%  &  0.036  \\
    \bottomrule
\end{tabular}
\end{table}

\section{Conclusion}
this paper introduces an innovative ensemble method, LLM-ensemble, designed to optimize product attribute value extraction by ensembling the outputs of various Large Language Models (LLMs). By dynamically learning the weights for different LLMs and aggregating their labels, our algorithm not only achieves theoretical optimality but also excels in efficiency, convergence speed, and deployment safety. Through comparison experiments with state-of-the-art LLMs on Walmart's internal dataset, the LLM-ensemble method has demonstrated superior performance over all individual LLMs. The A/B test of this method in production models has also improved the performance of our recommendation product.

\section{Presenter Bio}
Xiaohan Li is a Senior Data Scientist at the personalization team of Walmart Global Tech. He received his Ph.D. in Computer Science from the University of Illinois at Chicago (UIC) and B.Eng in Computer Science from Beijing University of Posts and Telecommunications (BUPT). His research interests are recommender systems, large language models, diffusion models, and graph neural networks.

\newpage

\bibliographystyle{ACM-Reference-Format}
\balance
\bibliography{sample-base}

\end{document}